\documentclass[review]{elsarticle}

\usepackage{lineno}
\modulolinenumbers[5]
\usepackage{balance}
\usepackage{comment}
\usepackage{multirow}
\usepackage{tikz,pgfplots,pgfplotstable,booktabs}
\usepackage{eurosym}
\usepackage[mathscr]{euscript}
\usepackage{amsmath,amssymb,amsfonts}
\usepackage{algorithmic}
\usepackage{graphicx}
\usepackage{textcomp}
\usepackage[utf8]{inputenc}
\usepackage{url}
\usepackage{bm}
\usepackage{hyperref}

\usepackage{tikz}
\usetikzlibrary{arrows,decorations.pathmorphing,backgrounds,fit,positioning,shapes.symbols,chains, shapes,snakes}

\newenvironment{ldescription}[1]
  {\begin{list}{}%
   {\renewcommand\makelabel[1]{##1\hfill}%
   \settowidth\labelwidth{\makelabel{#1}}%
   \setlength\leftmargin{\labelwidth}
   \addtolength\leftmargin{\labelsep}}}
  {\end{list}}

\journal{Journal of \LaTeX\ Templates}

\biboptions{authoryear}

\begin{document}

\begin{frontmatter}

\title{\textcolor{black}{Inverse Optimization with Kernel Regression: Application to the Power Forecasting and Bidding of a Fleet of Electric Vehicles}}

%\author{Ricardo Fernández-Blanco\fnref{a}, Juan Miguel Morales\fnref{a,b}, Salvador Pineda\fnref{a,c}, Álvaro Porras\fnref{a}}
%\address{University of Malaga, Malaga, Spain}
%\fntext[a]{OASYS research group}
%\fntext[b]{Department of Applied Mathematics}
%\fntext[c]{Department of Electrical Engineering}

\cortext[mycorrespondingauthor]{Corresponding author}

\author[a]{Ricardo Fernández-Blanco\corref{mycorrespondingauthor}}
\ead{Ricardo.FCarramolino@gmail.com}

\author[a,b]{Juan Miguel Morales}
\ead{Juan.Morales@uma.es}

\author[a,c]{Salvador Pineda}
\ead{spinedamorente@gmail.com}

\author[a]{Álvaro Porras}
\ead{alvaroporras19@gmail.com}

\address[a]{OASYS research group, University of Malaga, Malaga, Spain}
\address[b]{Department of Applied Mathematics, University of Malaga, Malaga, Spain}
\address[c]{Department of Electrical Engineering, University of Malaga, Malaga, Spain}

\begin{abstract}
This paper considers an aggregator of Electric Vehicles (EVs) who aims to learn the aggregate power of his/her fleet while also participating in the electricity market. The proposed approach is based on a data-driven inverse optimization (IO) method, which is highly nonlinear. To overcome such a caveat, we use a two-step estimation procedure which requires solving two convex programs. Both programs depend on penalty parameters that can be adjusted by using grid search. In addition, we propose the use of kernel regression to account for the nonlinear relationship between the behavior of the pool of EVs and the explanatory variables, i.e., the past electricity prices and EV fleet's driving patterns. Unlike any other forecasting method, the proposed IO framework also allows the aggregator to derive a bid/offer curve, i.e. the tuple of price-quantity to be submitted to the electricity market, according to the market rules. We show the benefits of the proposed method against the machine-learning techniques that are reported to exhibit the best forecasting performance for this application in the technical literature.
\end{abstract}


\end{frontmatter}

%\linenumbers

\section*{Nomenclature}
The main notation used throughout the text is stated below for quick reference. Other symbols are defined as required. 

\subsection*{Sets and Indices}
\begin{ldescription}{$xxxx$}
\item [$\mathcal{B}$] Set of energy blocks, indexed by $b$.
\item [$\mathcal{B}^{c/d}$] Set of energy blocks associated with the charging/discharging power, indexed by $b$.
%\item [$\mathcal{R}$] Set of regressors, indexed by $r$.
\item [$\mathcal{T}$] Set of time periods, indexed by $t$ and $\tau$.
\item [$\Omega^{X}$] Set of time periods belonging to the set $X = \{tr, v, test\}$ where $tr$, $v$, $test$ refer to the training, validation, and test set, in that order.
\end{ldescription}
\subsection*{Parameters}
\begin{ldescription}{$xxxxxxx$}
%\item [$\overline{P}_{b,t}$] Width of charging load block $b$ at time period $t$.
\item [$\underline{E}_{b,t}, \overline{E}_{b,t}$] Width for the aggregate discharging/charging power block $b$ in time period $t$ [kW].
\item [$H$] Feasibility penalty parameter.
\item [$K_{t,\tau}$] Value of the kernel on two feature vectors at time periods $t$ and $\tau$.
\item [$M$] Regularization hyper-parameter.
\item [$N_B$] Number of energy blocks.
\item [$\lambda_{t}$] Electricity price in time period $t$ [\euro/kWh].
%\item [$z_{r,t}$] Value of regressor $r$ in time period $t$.
\item [$\boldsymbol{z}_{t}$] Vector of regressors in period $t$.
\item [$\gamma$] Hyper-parameter related to the Gaussian kernel.
\end{ldescription}

\subsection*{Decision Variables}
\begin{ldescription}{$xxxxx$}
\item [$m_{b,t}$] Marginal utility of block $b$ of the aggregate power in time period $t$ [\euro/kWh].
\item [$p_{b,t}$] Power in block $b$ and time period $t$ [kW].
\item [$\underline{P}_t, \overline{P}_t$] Lower and upper bound for the aggregate power in time period $t$ [kW].
\item [$\underline{\alpha}_{t}, \overline{\alpha}_{t}$] Coefficient relative to the kernel regression of the lower/upper power bounds in period $t \in \Omega^{tr}$ [kW].
\item [$\epsilon_t$] Duality gap in time period $t \in \Omega^{tr}$ [\euro].
\item [$\underline{\mu}, \overline{\mu}$] Intercept for the lower/upper power bounds [kW].
\item [$\nu_b$] Intercept for the marginal utility of block $b$ [\euro/kWh].
\item [$\underline{\xi}^{+}_t$, $\underline{\xi}^{-}_t$] Slack variables associated with the lower power bound in time period $t$ [kW].
\item [$\overline{\xi}^{+}_t$, $\overline{\xi}^{-}_t$] Slack variables associated with the upper power bound in time period $t$ [kW].
\item [$\rho_{t}$] Coefficient relative to the kernel regression of the marginal utility in time period $t \in \Omega^{tr}$ [\euro/kWh].
\end{ldescription}

\section{Introduction}
According to the White Paper on transport of the \cite{COM2011_whitepaper}, one of the main goals to achieve a sustainable transport system is to \textit{halve the use of ‘conventionally fuelled’ cars in urban transport by 2030; phase them out in cities by 2050; achieve essentially CO$_2$-free city logistics in major urban centres by 2030}. This will spur the use of electric vehicles (EVs) across Europe \citep{COM2011_whitepaper}. Although nowadays the penetration of EVs in the European market is slow albeit steady, the estimated electricity demand from all EVs worldwide was 54 TWh in 2017 \citep{bunsen2018global}. Thus, the growing electrification of the road transport will impact the power system operation and planning of the future and, as a consequence, new actors and facilities will come into play, e.g. aggregator agents \citep{bandpey2018two}, or battery swap stations \citep{yang2015battery}. 

Within the context of the restructured power industry, the aggregator agents face several challenges: (i) the forecast of the charging power of the fleet of EVs in the short-term, and (ii) the determination of a bid curve to participate in the electricity market to maximize their profits when the fleet of EVs is large enough. \textcolor{black}{Nowadays, the EVs may be prepared with bi-directional vehicle-to-grid (V2G) capabilities, which means that the EVs can extract power from and inject power into the electrical grid while parked \citep{kempton2005vehicle}. This is possible as long as the EVs are equipped with the necessary smart metering-and-control infrastructure as well as a suitable connection to the electrical grid. In this case,} the aggregator will also need to forecast the EV-fleet discharging power. 

Short-term load forecasting is widely applied in the power sector to predict the electricity demand (and price) for different granularity levels \citep{shahidehpour2003market}. In the last years, EV charging load forecasting tools have been proposed in the technical literature by means of ARIMA-based models \citep{amini2016, korolko2015}; machine-learning techniques \citep{majidpour2016,sun2016,xydas2013}, such as support vector regression; or big data technologies \citep{arias2016}. All these papers neglected the bi-directional V2G capabilities of the EVs. Moreover, the above methodologies aimed to provide a single-purpose application, i.e., the forecasting of the charging power of either an EV or a fleet of EVs. Instead, we propose here a multi-purpose application for the aggregator of EVs in order to not only forecast the EV-fleet power, but also to derive a bid/offer curve according to the rules of the electricity market, e.g. see \cite{omie}.  

In this paper, we apply inverse optimization (IO) to forecast the EV-fleet power while deriving a bid/offer curve. The goal of an IO problem is to infer the optimization model parameters given a set of observed decision variables or measurements collected by an observer. For instance, \cite{zhang2010inverse} applied IO for linearly-constrained convex problems in the industrial and managerial areas but its application was limited to single observed decisions. \cite{aswani2018} proposed a statistically consistent methodology for IO when the measurements of the optimal decisions of a convex optimization problem are noisy. In a more general context, when the observer has imperfect information, \cite{esfahani2018data} devised a distributionally robust inverse optimization problem. IO has also been applied for equilibrium problems \citep{bertsimas2015data}, multiobjective convex optimization \citep{roland2016finding}, or robust optimization \citep{chan2019inverse}. However, few papers have implemented IO in the field of power systems \citep{saez2016data, Saez-Gallego2018,lu2018data, ruiz2013,zhou2010}. 

\cite{zhou2010} applied IO in the context of generation expansion planning to find an effective incentive policy; \cite{ruiz2013} estimated rival marginal offer prices for a strategic producer in a network-constrained day-ahead market by using IO; \cite{saez2016data} prescribed an IO approach by using bi-level programming to infer the market bid parameters of a pool of price-responsive consumers; in \cite{Saez-Gallego2018}, a novel IO approach was devised to statistically estimate the aggregate load of a pool of price-responsive buildings in the short-term; and, finally, \cite{lu2018data} applied IO to estimate the demand response characteristics of price-responsive consumers, as similarly done in \cite{saez2016data}. Unlike existing works \citep{saez2016data, Saez-Gallego2018,lu2018data, ruiz2013,zhou2010}, we address the EV-fleet power forecasting with an IO approach in which the prediction tool accounts for two distinctive features: (i) the pool of EVs may be equipped with V2G capabilities, and (ii) there may exist a strong nonlinear relationship between the EV-fleet power and the explanatory variables, namely past EVs' charging/discharging patterns and past electricity prices. To capture these nonlinear relations, we endogenously introduce kernels into the proposed IO approach. 

Kernels are widespread in the literature on machine learning, as can be seen in \cite{hofmann2008kernel,trevor2009elements,benitez2019cost}, just to name a few; and, in power systems, they were mainly used to predict electricity prices \citep{dudek2018probabilistic,kekatos2013day,kekatos2014electricity}. \cite{kekatos2013day} applied a kernel regression to forecast the electricity prices from the Midwest Independent System Operator day-ahead market in which the kernel itself is constructed by the product of three kernels: one for vectorial data and other two to account for non-vectorial data such as time and nodal information. This approach was generalized to low-rank kernel-based learning models in \cite{kekatos2014electricity}. Finally, \cite{dudek2018probabilistic} devised a probabilistic forecast method built on the Nadaraya-Watson estimator to predict the electricity prices from the Polish balancing and day-ahead markets.

The contributions of this paper are threefold: 

\begin{itemize}
    \item From a modeling perspective, we provide an IO framework to forecast the aggregate power of a fleet of EVs with V2G capabilities. In addition, the outcome of this framework may be used to bid/offer in the electricity market by using the estimated price-quantity tuples. To the best of the authors' knowledge, this is the first time in the technical literature that IO has been used to forecast the aggregate power of a price-responsive EVs' aggregator and to derive a suitable bid/offer curve for such an aggregator. 
    \item \textcolor{black}{We approximate the solution of the generalized IO problem by using a data-driven two-step estimation procedure. This procedure requires solving two different convex programming problems, which makes the process of building the forecasting model computationally affordable}. \textcolor{black}{A}s a salient feature of this work, a kernel is endogenously incorporated into the regression functions. 
    \item We thoroughly analyze the performance of the proposed methodology by using real-life data based on the latest National Household Travel Survey \citep{NHTS} and we compare the results against those provided by two machine-learning techniques, namely support vector regression and kernel-ridge regression. The former has been reported to exhibit the best forecasting performance for the present application in the technical literature \citep{xydas2013,sun2019optimal}.
\end{itemize}

The rest of the document is organized as follows: Section \ref{sec:methodology} provides the IO methodology; Section \ref{sec:benchmark} gives a general overview on the comparison methodologies; in Section \ref{sec:case}, we analyze a case study for a residential aggregator of EVs; conclusions are duly drawn in Section \ref{sec:conclusion}; and, finally, \ref{sec:simulator} presents a mixed-integer linear programming problem to generate synthetic data on the behavior of an EV fleet.  

\section{Inverse Optimization Methodology}
\label{sec:methodology}
To put the problem in context, we aim to forecast or learn the EV-fleet power $p_t$ (also known as aggregate power) in time period $t$ of \textcolor{black}{a price-responsive} aggregator, who is also interested in deriving a bid/offer curve to be submitted to the electricity market. The participants of the electricity market, namely consumers and producers, must submit a bid/offer curve consisting of blocks of energy and price. For the consumers, the bid curve should be monotonically non-increasing, whereas, for the producers, the offer curve should be monotonically non-decreasing, e.g. see \cite{omie}. \textcolor{black}{We assume a rational aggregator, which means that the market strategy of the EV fleet fundamentally relies on \emph{arbitrage}, by behaving as a consumer when the electricity price is low and, on the contrary, by acting as a producer when the price is high.} 

In order to predict the EV-fleet aggregate power and to derive a bid/offer curve, the aggregator may use past observed data, which are denoted as explanatory variables, features or regressors. \textcolor{black}{As one should expect for a price-responsive aggregator, t}he regressors in time period $t$ can be the lagged electricity price $\lambda^{\prime}_{t-l}$ or aggregate power $p^{\prime}_{t-l}$, $\forall l= 1, 2, ...$. In addition, past EV driving patterns, meteorological data, or categorical data (e.g., time information) can also be used for forecasting purposes.

Within this context, we first introduce the proposed forecasting\footnote{This problem is also known as forward or reconstruction problem in the IO jargon.} model in Section \ref{sec:forecasting_model}. Subsequently, Section \ref{sec:kernels} explains how we can account for past information. Finally, Section \ref{sec:estimation} thoroughly describes the two-step procedure to estimate the required parameters of the forecasting model. 

\subsection{Forward Model}
\label{sec:forecasting_model}
The key idea of this work is to forecast the EV-fleet power by using a simple optimization (linear programming) model which may, to some extent, \textit{mimic} its real behavior. In addition, unlike other forecasting techniques, this model is able to derive a bid/offer curve, as imposed by rules of electricity markets. Therefore, the formulation of the forward model that, we assume, represents the aggregate response of an EV fleet to the electricity prices at time period $t$, is mathematically expressed as: 
\begin{subequations}
\label{ev_agg}
\begin{align}
&\max_{p_{b,t}} \quad \sum_{b \in \mathcal{B}} p_{b,t} \left(m_{b,t} - \lambda_t \right) \label{fo_fwp} \\
& \text{subject to:} \notag\\
& \underline{P}_t \leq \sum_{b \in \mathcal{B}} p_{b,t} \leq \overline{P}_t : (\underline{\beta}_t, \overline{\beta}_t) \label{const1_fwp}  \\
& 0 \leq p_{b,t} \leq \overline{E}_{b,t} : (\underline{\phi}^c_{b,t}, \overline{\phi}^c_{b,t}), \quad \forall b \in \mathcal{B}^c \label{const2_fwp} \\
& \underline{E}_{b,t} \leq p_{b,t} \leq 0 : (\underline{\phi}^d_{b,t}, \overline{\phi}^d_{b,t}), \quad \forall b \in \mathcal{B}^d, \label{const3_fwp}
\end{align}
\end{subequations}
\noindent where dual variables are represented in parentheses after a colon in the respective constraints. For the sake of unit consistency, hourly time periods are considered. 

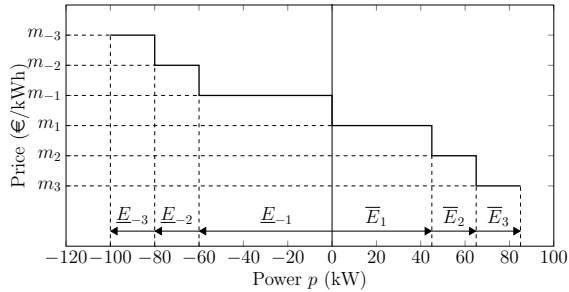
\begin{figure}[h] \centering
\begin{tikzpicture}[scale=0.5]
	\begin{axis}[	
    width=1.2\textwidth,
    height=8cm,
    xmin = -120,
    xmax = 100,
    ymin = 0,
    ymax = 80,
	clip marker paths=true,	
	xlabel = Power $p$ (kW),
	ylabel = Price (\euro/kWh),
	ytick = {20, 30, 40, 50, 60, 70},
	yticklabels={$m_3$, $m_2$, $m_1$, $m_{-1}$, $m_{-2}$, $m_{-3}$},
	label style={font=\Large},
    tick label style={font=\Large} ]
	\addplot[line width=1pt,draw=black] table [x=power, y=price, col sep=comma] {bid_example.csv}; %\addlegendentry{Data}
	\addplot[black] coordinates {(0,0)(0,100)};
	
	\addplot[dashed, black] coordinates {(-100,0)(-100,70)};
	\addplot[dashed, black] coordinates {(-80,0)(-80,60)};
	\addplot[dashed, black] coordinates {(-120,70)(-100,70)};
	\node[fill=white, font=\Large] at (30, 10) {$\underline{E}_{-3}$};
	%\node[fill=white, font=\Large] at (130, 70) {$m_{-3}$};
	\draw[<-,-triangle 60] (axis cs:-80, 5) -- (axis cs:-100, 5);
	
	\addplot[dashed, black] coordinates {(-60,0)(-60,50)};
	\addplot[dashed, black] coordinates {(-120,60)(-80,60)};
	\node[fill=white, font=\Large] at (50, 10) {$\underline{E}_{-2}$};
	%\node[fill=white, font=\Large] at (130, 60) {$m_{-2}$};
	\draw[<-,-triangle 60] (axis cs:-60, 5) -- (axis cs:-80, 5);
	
	\addplot[dashed, black] coordinates {(-120,50)(-60,50)};
	\node[fill=white, font=\Large] at (95, 10) {$\underline{E}_{-1}$};
	%\node[fill=white, font=\Large] at (130, 50) {$m_{-1}$};
	\draw[<-,-triangle 60] (axis cs: 0, 5) -- (axis cs:-60, 5);
	
	\addplot[dashed, black] coordinates {(45,0)(45,30)};
	\addplot[dashed, black] coordinates {(-120,40)(0,40)};
	\node[fill=white, font=\Large] at (140, 10) {$\overline{E}_{1}$};
	%\node[fill=white, font=\Large] at (113, 40) {$m_{1}$};
	\draw[->,-triangle 60] (axis cs: 0, 5) -- (axis cs:45, 5);
	
	\addplot[dashed, black] coordinates {(65,0)(65,20)};
	\addplot[dashed, black] coordinates {(-120,30)(45,30)};
	\node[fill=white, font=\Large] at (175, 10) {$\overline{E}_{2}$};
	%\node[fill=white, font=\Large] at (113, 30) {$m_{2}$};
	\draw[->,-triangle 60] (axis cs: 45, 5) -- (axis cs:65, 5);
	
	\addplot[dashed, black] coordinates {(85,0)(85,20)};
	\addplot[dashed, black] coordinates {(-120,20)(65,20)};
	\node[fill=white, font=\Large] at (195, 10) {$\overline{E}_{3}$};
	%\node[fill=white, font=\Large] at (113, 20) {$m_{3}$};
	\draw[->,-triangle 60] (axis cs: 65, 5) -- (axis cs:85, 5);
	\end{axis}	
\end{tikzpicture} \\
\vspace{-0.2cm}
\caption{Three-block stepwise offer (bid) price function of the EVs' aggregator. In this example, the offer (bid) price function is represented to the left (right) of the y-axis, and the sets $\mathcal{B}^d = \{-3, -2, -1\}$ and $\mathcal{B}^c = \{1, 2, 3\}$. } \label{fig:bid}
\end{figure}

The reconstruction problem \eqref{ev_agg} aims to maximize the welfare of the EV aggregator, as given by the objective function \eqref{fo_fwp}. This objective function is made up of the EV fleet's surplus, which is related to the aggregate charging \textcolor{black}{and discharging} power. \textcolor{black}{The aggregate power is positive when the EVs' aggregator is charging, i.e., it behaves as a consumer. Otherwise, the aggregate power takes on negative values when the aggregator is discharging, i.e., it acts as a producer}. We assume step-wise offer/bid price functions as depicted in Fig. \ref{fig:bid}, as is customary in real-world electricity markets, e.g. see \cite{omie}. Constraints \eqref{const1_fwp} represent the lower and upper bounds on the aggregate power. Constraints \eqref{const2_fwp} impose the lower and upper bound on each block $b$ within the set $\mathcal{B}^c$ of charging power blocks. Since the charging power is assumed to be non-negative, then $p_{b,t}$ is bounded between $0$ and a positive power bound $\overline{E}_{b,t}$. Likewise, constraints \eqref{const3_fwp} impose the lower and upper bound on each block $b$ within the set $\mathcal{B}^d$ of the discharging power blocks. We assume that the discharging power is non-positive and thus $p_{b,t}$ is bounded between a negative power bound $\underline{E}_{b,t}$ and $0$. Note that the total power $p_t = \sum_{b} p_{b,t}$. 

As previously stated, we want to {\color{black} anticipate} the EV-fleet power response by solving \eqref{ev_agg}. However, to this end, the set of parameters $\Phi = \{ \underline{E}_{b,t}, \overline{E}_{b,t}, m_{b,t}, \underline{P}_t, $ $ \overline{P}_t \}$ needs to be estimated since they are a priori unknown. \textcolor{black}{These parameters should be functions of time and of any regressor that the forecaster may consider meaningful and explanatory of the EV-fleet's operational behavior and, therefore, are to be inferred from past observations of the} aggregate power $p^{\prime}_t$, \textcolor{black}{the} electricity price $\lambda^{\prime}_t$, and  \textcolor{black}{the regressors that are eventually considered}. This fact gives rise to a generalized IO problem, which is highly nonlinear and non-convex. This problem can be naturally formulated as a bilevel optimization problem, which may be computationally nonviable when moderately increasing the sample size. To deal with such complexity, we apply a methodology that builds on the one first proposed in \cite{Saez-Gallego2018}. In that paper, however, the regression function is linear in their features and may be limited to capture nonlinear relations between the EV-fleet power and the regressors. To circumvent such a caveat, and as one of the salient features of this work, we incorporate kernels into the regression functions. Furthermore, the forward model we propose, i.e. problem \eqref{ev_agg}, allows for power intakes and outputs, unlike the one used in \cite{Saez-Gallego2018}. This extra dose of model flexibility is critical to capture the behavior of an EV fleet with V2G capabilities \textcolor{black}{since the aggregator power may be positive when the net power comes from the grid (i.e. the aggregator acts as a consumer) or negative when the net power flows into the grid (i.e. the aggregator behaves as a producer). Therefore, the forecasting model is tailored to account for this dual operational mode by introducing differentiated marginal utility blocks both for charging and discharging.} 

\subsection{Accounting for Past Information: Kernels}
\label{sec:kernels}
In the realm of machine learning, the kernel functions are rather popular in learning algorithms \citep{hofmann2008kernel} since they are able to capture nonlinear relationships between the dependent and the explanatory variables. Unlike in \cite{Saez-Gallego2018}, where affine functions were used to model the dependence of the parameters of the forward model \eqref{ev_agg} on the regressors, we propose the use of kernel regressions to estimate $\underline{P}_t$, $\overline{P}_t$, and $m_{b,t}$:
\begin{align}
&  \underline{P}_t = \underline{\mu} + \sum_{\tau \in \Omega^{tr}} \underline{\alpha}_{\tau} K_{t,\tau}, \quad \forall t \in \mathcal{T} \label{pmin_kernel_regression} \\
& \overline{P}_t = \overline{\mu} + \sum_{\tau \in \Omega^{tr}} \overline{\alpha}_{\tau} K_{t,\tau}, \quad \forall t \in \mathcal{T} \label{pmax_kernel_regression}\\
& m_{b,t} = \nu_b + \sum_{\tau \in \Omega^{tr}} \rho_{\tau} K_{t,\tau}, \quad \forall t \in \mathcal{T}. \label{m_kernel_regression}
\end{align}

Many kernel functions can be used: polynomial, hyperbolic tangent, Gaussian, among others. For the sake of illustration purposes, the Gaussian kernel \citep{trevor2009elements} can be defined as follows:
\begin{align}
& K_{t,\tau} = K \left( \boldsymbol{z}_t, \boldsymbol{z}_{\tau}\right) = e^{-\gamma \lVert \boldsymbol{z}_t - \boldsymbol{z}_{\tau} \rVert_2^2}, \quad \forall t \in \mathcal{T}, \tau \in \Omega^{tr}, \label{kernel_eq1}
\end{align}

\noindent wherein $\gamma$ is a scale parameter inversely proportional to the variance of the Gaussian function; and $\lVert \boldsymbol{z}_t - \boldsymbol{z}_{\tau} \rVert_2^2$ is the squared Euclidean distance between two feature vectors at time periods $t$ and $\tau$. Thus, the Gaussian kernel can be interpreted as a similarity measure between two time periods, i.e., if the two feature vectors are identical $\boldsymbol{z}_t=\boldsymbol{z}_{\tau}$, then the value of $K_{t,\tau} = 1$, otherwise its value ranges in the interval $(0, 1]$. 

\textcolor{black}{As previously mentioned, meaningful or explanatory features should be used in the kernel regression function for adequately inferring the estimates. In the proposed IO methodology, the power bounds $\underline{P}_t$ and $\overline{P}_t$ are key to capturing the price-responsiveness of the aggregator since they determine the width and the range of the step-wise price-response function of the EV fleet. For instance, if the electricity price is high, one should expect that the EV fleet will behave as a producer (in V2G mode) and therefore, the power bounds would take on negative values, this way producing a step-wise \emph{offer} curve displaced towards negative power values (i.e., discharge). On the contrary, if the price is low, one should expect the opposite: the EV fleet would act as a consumer, with the power bounds taking positive values and defining a step-wise \emph{bidding} curve displaced towards positive power values (i.e., charge). On the other hand, the marginal utilities aim to capture the price-sensitivity of the EVs aggregate power. Therefore, by making both the power bounds and the marginal utilities dependent on past prices along with past values of the aggregate EV-fleet power and/or other external factors, we can capture the changes in the EV-fleet power due to price variations over time.}

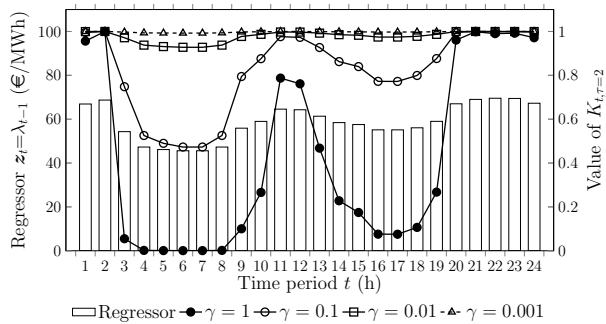
\begin{figure}[h] \centering
\begin{tikzpicture}[scale=0.5]
	\begin{axis}[
    width=1.2\textwidth,
    height=8cm,
    xmin = 0.5,
    xmax = 25.5,
    ymin = 0,
    ymax = 110,
	%legend style={at={(0.5,0.96)},anchor=north,legend cell align=left,legend columns=2},
	%clip marker paths=true,	
	ylabel = Regressor $\boldsymbol{z}_t \text{=} \lambda_{t-1}$ (\euro/MWh),
	xlabel = Time period $t$ (h),
	xtick= {1,2,3,4,5,6,7,8,9,10,11,12,13,14,15,16,17,18,19,20,21,22,23,24},
	ytick={0, 20, 40, 60, 80, 100},
	major grid style={line width=0,draw=white!50},
	label style={font=\Large},
    tick label style={font=\large},
    ybar interval=0.6]
	\addplot[area legend] table [x=t, y=regressor, col sep=comma] {kernel_plot_data_with_regressors.csv};\label{plot_one_k} 
	
	\end{axis}
	
	\begin{axis}[
	ylabel near ticks, 
	yticklabel pos=right,
    axis x line=none,
    width=1.2\textwidth,
    height=8cm,
    xmin = 0,
    xmax = 25,
    ymin = 0,
    ymax = 1.1,
	%legend style={at={(0.5,0.96)},anchor=north,legend cell align=left,legend columns=2},
	%clip marker paths=true,	
	legend style={at={(0.5,-0.2)},anchor=north, legend columns=5, draw=none,font=\Large},
	ylabel = Value of $K_{t,\tau=2}$,
	xtick= {1,2,3,4,5,6,7,8,9,10,11,12,13,14,15,16,17,18,19,20,21,22,23,24},
	ytick={0, 0.2, 0.4, 0.6, 0.8, 1.0},
	label style={font=\Large},
    tick label style={font=\large}]
    \addlegendimage{/pgfplots/refstyle=plot_one_k}\addlegendentry{Regressor}
	\addplot[line width=1pt, mark=*, mark size=3,  mark options={solid}, draw=black] table [x=t, y=1, col sep=comma] {kernel_plot_data_with_regressors.csv}; \label{plot_two_k} \addlegendentry{$\gamma=1$}
	\addplot[line width=1pt, mark=o, mark size=3, mark options={solid}, draw=black] table [x=t, y=0.1, col sep=comma] {kernel_plot_data_with_regressors.csv};\label{plot_three_k} \addlegendentry{$\gamma=0.1$}
	\addplot[line width=1pt, mark=square, mark size=3, mark options={solid}, draw=black] table [x=t, y=0.01, col sep=comma] {kernel_plot_data_with_regressors.csv};\label{plot_three_k} \addlegendentry{$\gamma=0.01$}
    \addplot[dashed, line width=1pt, mark=triangle, mark size=3, mark options={solid}, draw=black] table [x=t, y=0.001, col sep=comma] {kernel_plot_data_with_regressors.csv};\label{plot_three_k} \addlegendentry{$\gamma=0.001$}
	
	\end{axis}

\end{tikzpicture} \\
\vspace{-0.2cm}
\caption{Values of the Gaussian kernel for each time period $t$ of a day with respect to period $\tau = 2$, i.e., $K_{t, \tau=2}$, for different values of the parameter $\gamma$ in the right y-axis and the corresponding regressor values in the left y-axis. } \label{fig:kernel}
\end{figure}

\textcolor{black}{\emph{Illustrative example.} L}et us assume that $\boldsymbol{z}_t$ comprises only one regressor, {\color{black} e.g.} the electricity price in the previous time period, i.e., $\boldsymbol{z}_t$ = $\lambda_{t-1}$. Thus, Fig. \ref{fig:kernel} provides the values of the kernel for each time period $t$ of a day with respect to the second time period $\tau = 2$, i.e., $K_{t, \tau=2}$, for different values of parameter $\gamma$. Moreover, the values of the regressor $\boldsymbol{z}_t$ for the 24 hours are shown in the figure. We can observe that high values of $\gamma$ lead to kernel values equal to $1$ just when the two regressors are very close to each other (e.g., see time periods 21--23 for $\gamma=1$); conversely, low values of $\gamma$ lead to kernel values equal to $1$ even when the regressors are very different from each other (e.g., see values for all time periods when $\gamma=0.001$). Therefore, we need to carefully tune the hyper-parameter $\gamma$, as described in Section \ref{tuning}.

\subsection{Two-step Estimation Procedure}
\label{sec:estimation}
The thrust of this work is the estimation of the set of parameters $\Phi = \{ \underline{E}_{b,t}, \overline{E}_{b,t}, m_{b,t}, \underline{P}_t, \overline{P_t} \}$ and the corresponding coefficient estimates $\underline{\mu}, \underline{\alpha}_t, \overline{\mu}, \overline{\alpha_t}, $ $ \nu_b, \rho_t$ of the regression functions described in \eqref{pmin_kernel_regression}--\eqref{m_kernel_regression}. To do that, we can use bilevel optimization, however, as mentioned previously, it may lead to a prohibitive computational burden when moderately increasing the sample size. Therefore, we resort to a two-step procedure based on two convex programming problems: (i) the \emph{feasibility problem}, which is devoted to estimating all parameters that determine the feasibility of the observed EV-fleet power values in the forward problem \eqref{ev_agg} (i.e., the power bounds), and (ii) the \emph{optimality problem}, which estimates the marginal utility of the EVs' aggregator, i.e., the parameters of problem \eqref{ev_agg} that are related to the optimality of the observed power values. The key idea of the \emph{feasibility problem} is to shape the power bounds $\underline{P}_{t}$ and $\overline{P}_{t}$ so that a certain percentage $H$ of the observed EV-fleet power values are feasible for the forward problem \eqref{ev_agg}. Note that the width for the aggregate power blocks $\underline{E}_{b,t}$ and $\overline{E}_{b,t}$ can be easily computed from the estimated power bounds by assuming that the energy blocks are all of same length. Conversely, the \emph{optimality problem} estimates the marginal utilities $m_{b,t}$ driven by the minimization of the duality gap of the forward problem once the power bounds are fixed. Its aim is thus to make the observed EV-fleet power values as optimal as possible for problem \eqref{ev_agg} (recall that we use \eqref{ev_agg} as the forward model). It should be noted that the pair $(m_{b,t},\overline{E}_{b,t})$ for all blocks constitutes the bid curve of the aggregator at time period $t$. Likewise, the pair $(m_{b,t},-\underline{E}_{b,t})$ for all blocks constitutes the offer curve of the aggregator at time period $t$. In practice, those curves may be submitted to the market operator, who is the entity responsible for the financial management of electricity markets, e.g. see \cite{omie}.

\subsubsection{Feasibility Problem}
\label{sec:feas_problem}
Given a fixed value of control parameter $H \in [0, 1)$, this problem can be formulated as:
\begin{subequations}
\label{feasibility_problem}
\begin{align}
&\min_{\Xi^{fp}} \sum_{t \in \Omega^{tr}}  H \left( \overline{\xi}_t^{-}  + \underline{\xi}_t^{-} \right)  + \sum_{t \in \Omega^{tr}} \left( 1 - H \right) \left( \overline{\xi}_t^{+} + \underline{\xi}_t^{+} \right)  \label{fo_fp}\\
& \text{subject to:} \notag\\
& \overline{P}_t - p^{\prime}_t = \overline{\xi}_t^{+} - \overline{\xi}_t^{-}, \quad \forall t \in \Omega^{tr} \label{const1_fp}\\
& p^{\prime}_t - \underline{P}_t = \underline{\xi}_t^{+} - \underline{\xi}_t^{-}, \quad \forall t \in \Omega^{tr} \label{const2_fp}\\
& \overline{P}_t \geq \underline{P}_t, \quad \forall t \in \Omega^{tr} \label{const3_fp}\\
& \text{Constraints \eqref{pmin_kernel_regression}--\eqref{pmax_kernel_regression}} \label{const4_fp}\\
& \overline{\xi}_t^{+}, \underline{\xi}_t^{+}, \overline{\xi}_t^{-}, \underline{\xi}_t^{-} \geq 0, \quad \forall t \in \Omega^{tr}, \label{const5_fp}
\end{align}
\end{subequations}  

\noindent where the set of variables to be optimized is $\Xi^{fp} = \{ \underline{P}_t, \overline{P}_t,$ $ \overline{\xi}_t^{+}, \underline{\xi}_t^{+}, \overline{\xi}_t^{-}, \underline{\xi}_t^{-}, \underline{\mu},$ $\overline{\mu}, \underline{\alpha}_t, \overline{\alpha}_t\}$. Note that problem \eqref{feasibility_problem} is a convex program.

The objective function \eqref{fo_fp} minimizes the sum of feasibility and infeasibility slack variables associated with the power bounds. Constraints \eqref{const1_fp}--\eqref{const2_fp} are the power bound constraints with the feasibility and infeasibility slack variables, where $p^{\prime}_t$ is the observed EV-fleet power value at time period $t$. Constraints \eqref{const3_fp} ensure that the upper bound of the aggregate power is greater than its respective lower bound. Constraints \eqref{const4_fp} impose kernel regression functions for the power bounds wherein the coefficients to be estimated are $\underline{\mu}$,  $\overline{\mu}$, $\underline{\alpha}_t$, $\overline{\alpha}_t$. Finally, constraints \eqref{const5_fp} declare the variables $\overline{\xi}_t^{+}, \underline{\xi}_t^{+}, \overline{\xi}_t^{-}, \underline{\xi}_t^{-}$ as non-negative. Importantly, the higher the value of $H$, the wider the power bounds delivered by \eqref{feasibility_problem} and, therefore, the more price-responsive the EV fleet is expected to be.   

The use of kernels increases the flexibility of the regression function when increasing the size of the training set. However, it also tends to over-fitting. To control the risk of over-fitting, a regularization parameter $M \in [0, 1]$ is used to factor in the sum of the squared values of the coefficient estimates $\underline{\alpha}_{t}$ and $\overline{\alpha}_{t}$, similarly to what is typically done in kernel-ridge regression \citep{trevor2009elements}. Thus, the objective function \eqref{fo_fp} \textcolor{black}{should be replaced with}:
{\color{black}
\begin{align}
&\min_{\Xi^{fp}} M \sum_{t \in \Omega^{tr}} \left( \underline{\alpha}_{t}^2 + \overline{\alpha}_{t}^2 \right) \notag\\
& \hspace{2cm}+ \left( 1 - M \right) \bigl[\sum_{t \in \Omega^{tr}}  H \left( \overline{\xi}_t^{-}  + \underline{\xi}_t^{-} \right)  + \sum_{t \in \Omega^{tr}} \left( 1 - H \right) \left( \overline{\xi}_t^{+} + \underline{\xi}_t^{+} \right)\bigr].
\end{align}
Both hyper-parameters $M$ and $H$ in the objective function and parameter $\gamma$ of the kernel regression function must be adequately adjusted to modulate the power bounds to the observed EV-fleet power values so that the out-of-sample forecasting error is minimized.}

\subsubsection{Optimality Problem}
\label{sec:opt_problem}
Once the power bounds (i.e., $\widehat{\underline{P}}_t$, $\widehat{\overline{P}}_t$) are estimated from \eqref{feasibility_problem}, we can compute the power block limits $\widehat{\overline{E}}_{b,t}$, $\forall b \in \mathcal{B}^c$ and $\widehat{\underline{E}}_{b,t}$, $\forall b \in \mathcal{B}^d$ based on the assignments described in Table \ref{tab:power_block_limits}. The optimality problem can then be derived by using results from duality theory of linear programming and it can be formulated as: 
%
%It is worth mentioning that the aggregated power needs to be adjusted according to the infeasibility slack variables. For instance,  $\widetilde{p}^{\prime}_t = p^{\prime}_t - \widehat{\overline{\xi}}_t^{-} + \widehat{\underline{\xi}}_t^{-}$. Thus, the \emph{optimality} problem can be formulated as:
%
\begin{subequations}
\label{optimality_problem}
\begin{align}
&\min_{\Xi^{op}} \quad \sum_{t \in \Omega^{tr}} \epsilon_t \label{fo_op}\\
%& \text{subject to:} \notag\\
& \widehat{\overline{P}}_t \overline{\beta}_t - \widehat{\underline{P}}_t \underline{\beta}_t + \sum_{b \in \mathcal{B}^c} \widehat{\overline{E}}_{b,t} \overline{\phi}^c_{b,t} - \sum_{b \in \mathcal{B}^d} \widehat{\underline{E}}_{b,t} \underline{\phi}^d_{b,t} - \epsilon_t = \notag\\
&\hspace{6.5cm}\sum_{b \in \mathcal{B}} p^{\prime}_{b,t} \left( m_{b,t} - \lambda_t\right), \forall t \in \Omega^{tr} \label{const1_op}\\
& - \underline{\phi}^c_{b,t} + \overline{\phi}^c_{b,t} - \underline{\beta}_t + \overline{\beta}_t = m_{b,t} - \lambda_t,  \quad \forall b \in \mathcal{B}^c, t \in \Omega^{tr} \label{const2_op}\\
& - \underline{\phi}^d_{b,t} + \overline{\phi}^d_{b,t} - \underline{\beta}_t + \overline{\beta}_t = m_{b,t} - \lambda_t,  \quad \forall b \in \mathcal{B}^d, t \in \Omega^{tr} \label{const3_op}\\
&\text{Constraints \eqref{m_kernel_regression}} \label{const4_op}\\
& \nu_b \geq \nu_{b+1}, \quad \forall b \in \mathcal{B} \setminus \{b = N_B\} \label{const5_op}\\ 
%& \epsilon_t \geq 0 \quad \text{if} \quad p^{\prime}_t \in [\widehat{\underline{P}}_t, \widehat{\overline{P}}_t], \quad \forall t \in \Omega^{tr}\label{const6_op}\\
&  \underline{\beta}_t, \overline{\beta}_t, \underline{\phi}^c_{b,t}, \overline{\phi}^c_{b,t}, \underline{\phi}^d_{b,t}, \overline{\phi}^d_{b,t} \geq 0, \quad \forall t \in \Omega^{tr}, \label{const6_op}
\end{align}
\end{subequations} 

\noindent where the set of decision variables is $\Xi^{op} = \{ m_{b,t}, \epsilon_t,\underline{\beta}_t, \overline{\beta}_t, \underline{\phi}^c_{b,t}, \overline{\phi}^c_{b,t}, \underline{\phi}^d_{b,t}, \overline{\phi}^d_{b,t}, $ $ \nu_b, \rho_t \}$. Note that problem \eqref{optimality_problem} is a convex program. 

\begin{table}[h!]
\caption{Value of $\widehat{\overline{E}}_{b,t}$, $\forall b \in \mathcal{B}^c$ and $\widehat{\underline{E}}_{b,t}$, $\forall b \in \mathcal{B}^d$}
\label{tab:power_block_limits}
\centering
\begin{tabular}{c@{\hspace{1\tabcolsep}}c@{\hspace{1\tabcolsep}}c@{\hspace{1\tabcolsep}}c@{\hspace{1\tabcolsep}}c}
\cline{3-5}
\\[-17pt]
\multicolumn{2}{c}{} &  $\widehat{\overline{P}}_t \geq \widehat{\underline{P}}_t \geq 0$ &  $\widehat{\underline{P}}_t \leq \widehat{\overline{P}}_t \leq 0$  &  $\widehat{\overline{P}}_t \geq 0 \geq \widehat{\underline{P}}_t$ \\
\hline
\multirow{2}{*}{$\widehat{\overline{E}}_{b,t}$} & $b=1$  & $\widehat{\underline{P}}_t$ & $0$ & $\widehat{\overline{P}}_t/N_B$ \\
                                      & $b \in \mathcal{B}^c \setminus \{1\}$  & $\frac{\left(\widehat{\overline{P}}_t - \widehat{\underline{P}}_t\right)}{N_B - 1}$ & $0$ & $\widehat{\overline{P}}_t/N_B$ \\
\hline
\multirow{2}{*}{$\widehat{\underline{E}}_{b,t}$} & $b=-1$  & $0$ & $\widehat{\overline{P}}_t$ & $\widehat{\underline{P}}_t/N_B$ \\
                                       & $b \in \mathcal{B}^d \setminus \{-1\}$  & $0$ & $\frac{\left(\widehat{\underline{P}}_t - \widehat{\overline{P}}_t\right)}{N_B - 1}$ & $\widehat{\underline{P}}_t/N_B$ \\
\hline
\end{tabular}
\end{table}

The objective function \eqref{fo_op} minimizes the sum of the duality gaps of problem \eqref{ev_agg}. 
%when $M^{op} = 0$; otherwise, parameter $M^{op} \in (0, 1)$ controls the risk of over-fitting by weighting the sum of the squared values of the coefficient estimates $\alpha_t^m$. 
Constraints \eqref{const1_op} is the relaxed equality constraint associated with the strong duality theorem. Constraints \eqref{const2_op}--\eqref{const3_op}, \eqref{const6_op} are the dual feasibility constraints. Constraints \eqref{const4_op} impose a kernel regression function, with $\nu_b$ and $\rho_t$ as the coefficients to be estimated, in order to relate the marginal utilities and the regressors. Finally, constraints \eqref{const5_op} set the marginal utilities to be monotonically non-increasing, as imposed by rules in electricity markets \citep{omie}.

\subsubsection{Statistical Computation of Hyper-Parameters}
\label{tuning}
The main goal of this work is to learn the EV-fleet power for each period $t \in \Omega^{test}$ with the forward model \eqref{ev_agg}, which relies on the knowledge of a series of parameters, i.e., the power bounds and the marginal utilities. Those parameters are estimated with the models described in Sections \ref{sec:feas_problem} and \ref{sec:opt_problem}, whose outcome depend on the value of three hyper-parameters: $H$, $M$, and $\gamma$. Their optimal values are computed by using a grid search technique. We recursively solve problems \eqref{feasibility_problem} and \eqref{optimality_problem} for the training set $\Omega^{tr}$; and we then solve the forward problem \eqref{ev_agg} over the validation set $\Omega^{v}$ by using the estimated parameters $\Phi = \{ \underline{E}_{b,t}, \overline{E}_{b,t}, m_{b,t}, \underline{P}_t, \overline{P}_t \}$ as well as the electricity price at time period $t \in \Omega^{v}$. Thus, we set as the optimal values of the hyper-parameters those that lead to the least out-of-sample forecasting error in $\Omega^v$. 

\vspace{-0.3cm}
\section{Comparison Methodologies}
\label{sec:benchmark}
We compare the performance of the proposed kernel-based IO approach, hereinafter referred to as \textit{kio}, against (i) the state-of-the-art model to forecast the EV-fleet power, namely kernelized support vector regression (\textit{svr}), (ii) a kernel-ridge regression model (\textit{krr}), (iii) an IO approach with linear kernels (\textit{lio}), and (iv) persistence or naive models. Note that we use a Gaussian kernel in the regression functions of the \textit{feasibility problem} and a linear kernel in the regression function of the \textit{optimality problem}, as this combination exhibited the best trade-off between forecasting performance and simplicity in our numerical experiments.

Regarding the \textit{svr} and \textit{krr}, we respectively use the epsilon-\textit{svr} and the kernel-ridge regression models implemented in the scikit-learn library \citep{scikit-learn} under the Python programming language. The interested reader is referred to \cite{smola2004tutorial} for a detailed description of the \textit{svr}. For the sake of comparison, we also use the Gaussian kernel and we tune the corresponding hyper-parameters via grid search. Specifically, we tune the cost of constraints violation $C$ and the parameter associated with the kernel $\gamma$ for \textit{svr}; and the penalty parameter $\delta$ and the $\gamma$ parameter for \textit{krr}.

Regarding the naive models, we use three different ones since the EV-fleet power may experience seasonal patterns: \textit{h-naive}, \textit{d-naive}, and \textit{w-naive}, in which the forecast value of the aggregate power at time $t$ is equal to the observed value at time $t-1$, $t-24$, and $t-168$, in that order. Note that the forecast error of the naive models provides insight into the difficulty of prediction. 

The performance of the methods is compared with two metrics: the mean absolute error (MAE) and the root mean square error (RMSE) on the test set.

\section{Case Study}
\label{sec:case}
We first describe the data used for the case study in Section \ref{sec:ev_fleet_data}. Subsequently, we comprehensively analyze the results from the proposed approach for three cases of charging behavior without enabling the V2G capabilities in Section \ref{sec:results_g2v}. Finally, Section \ref{sec:results_v2g} presents the results for two cases when the electric vehicles are integrated with V2G services.

\subsection{EV-fleet Data}
\label{sec:ev_fleet_data}
For learning purposes, we would be only interested in the time series of electricity prices, the aggregate power of an EV fleet, and the total number of available vehicles to charge or discharge. However, to our knowledge, there is no real-life data available about an EVs' aggregator. Thus, we resort to the formulation of an optimization problem to simulate the behavior of such an EV fleet. The interested reader is referred to \ref{sec:simulator} for a detailed description of this simulator.

We assume a residential aggregator with 100 EVs. For the sake of simplicity, the technical parameters associated with each EV are identical: The maximum charging rate is 7.4 kW, the round-trip efficiency is 0.95, the minimum and maximum energy rates are 10 and 51 kWh, in that order, and the energy rating per kilometer is 0.137 kWh/km \citep{Technical_ZOE}. Due to the lack of real-life data about the parameters associated with the driving patterns (availability profiles and energy required for transportation) of EVs, we resort to the National Household Travel Survey \citep{NHTS}. From this data base, we can extract the availability status by using the departure/arrival time periods for each daily trip. Specifically, we assume that the EV is available until it begins its first daily trip and after it returns from its last daily trip for each day of the year. Otherwise the EV is unavailable and thus it may be in a motion status. The energy required for transportation $\chi_{v,t}$ can be computed as the product of the travelled distance and energy rating per kilometer (i.e., 0.137 kWh/km).

The electricity prices are obtained from the ENTSO-e Transparency Platform \citep{ENTSOE} for the period comprising January 9$^{th}$ till February 19$^{th}$ in Spain. We also assume that the load shedding cost $C^P = 1000$ $\textup{\euro}$/kWh. We run daily simulations with 15-min time steps to build a synthetic database for a pool of EVs. 

The simulations have been performed on a Linux-based server with one CPU clocking at 2.6 GHz and 2 GB of RAM using CPLEX 12.6.3 \citep{Cplex} under Pyomo 5.2 \citep{Pyomo}. Optimality gap is set to 0\%. \textcolor{black}{The input data files for reproducing the results have been shared with the scientific community in \url{https://github.com/groupoasys/Aggregated-EV-data}}.

\subsection{Forecast Results without Enabling V2G Capabilities}
\label{sec:results_g2v}
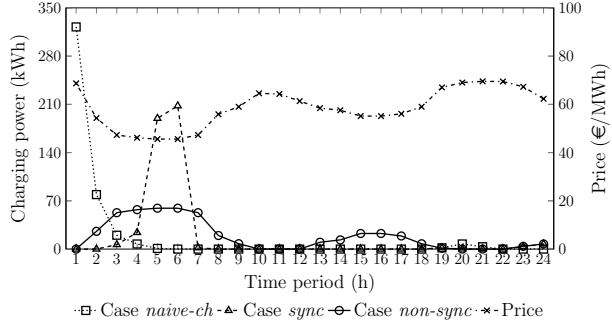
\begin{figure}[h] \centering
\begin{tikzpicture}[scale=0.5]
	\begin{axis}[
    width=1.2\textwidth,
    height=8cm,
    xmin = 0.5,
    xmax = 24.5,
    ymin = 0,
    ymax = 350,
	%legend style={at={(0.5,0.96)},anchor=north,legend cell align=left,legend columns=2},
	%clip marker paths=true,	
	ylabel = Charging power (kWh),
	xlabel = Time period (h),
	xtick= {1,2,3,4,5,6,7,8,9,10,11,12,13,14,15,16,17,18,19,20,21,22,23,24},
	ytick={0, 70, 140, 210, 280, 350},
	major grid style={line width=0,draw=white!50},
	label style={font=\Large},
    tick label style={font=\large}]
	\addplot[dashed, line width=1pt, mark=triangle, mark size=3, mark options={solid}, draw=black] table [x=t, y=g2v_a0, col sep=comma] {power_day1_tr_g2v.csv};\label{plot_one} 
	\addplot[line width=1pt, mark=o, mark size=3, mark options={solid}, draw=black] table [x=t, y=g2v_anot0, col sep=comma] {power_day1_tr_g2v.csv};\label{plot_two}
	\addplot[dotted, line width=1pt, mark=square, mark size=3, mark options={solid}, draw=black] table [x=t, y=g2v_naive, col sep=comma] {power_day1_tr_g2v.csv};\label{plot_three}
	
	\end{axis}
	
	\begin{axis}[
	ylabel near ticks, 
	yticklabel pos=right,
    axis x line=none,
    width=1.2\textwidth,
    height=8cm,
    xmin = 0.5,
    xmax = 24.5,
    ymin = 0,
    ymax = 100,
	%legend style={at={(0.5,0.96)},anchor=north,legend cell align=left,legend columns=2},
	%clip marker paths=true,	
	legend style={at={(0.5,-0.2)},anchor=north, legend columns=5, draw=none,font=\Large},
	ylabel = Price (\euro/MWh),
	xtick= {1,2,3,4,5,6,7,8,9,10,11,12,13,14,15,16,17,18,19,20,21,22,23,24},
	ytick={0, 20, 40, 60, 80, 100},
	label style={font=\Large},
    tick label style={font=\large}]
    \addlegendimage{/pgfplots/refstyle=plot_three}\addlegendentry{Case \textit{naive-ch}}
    \addlegendimage{/pgfplots/refstyle=plot_one}\addlegendentry{Case \textit{sync}}
    \addlegendimage{/pgfplots/refstyle=plot_two}\addlegendentry{Case \textit{non-sync}}
    \addplot[dash pattern=on 1pt off 3pt on 3pt off 3pt, line width=1pt, mark=x, mark size=3, mark options={solid}, draw=black] table [x=t, y=price, col sep=comma] {power_day1_tr_g2v.csv};\label{plot_four}  \addlegendentry{Price}
	
	\end{axis}

\end{tikzpicture} \\
\vspace{-0.3cm}
\caption{Charging power for cases \textit{naive-ch}, \textit{sync}, and \textit{non-sync} in the left y-axis and the corresponding electricity prices in the right y-axis.} \label{fig:power_sync}
\end{figure}

We assume that EVs do not enable their V2G capabilities (i.e. $B^d_v = 0$ in the model \eqref{sim_of}--\eqref{sim_eq8} in the \ref{sec:simulator}) and we compare the results for three cases: (i) a case in which the EVs satisfy their energy needs by using a \textit{naive} charging; (ii) a case in which the charging is highly synchronized, which occurs when $C^S$ is set to $0$ in \eqref{sim_of}--\eqref{sim_eq8}; and (iii) a case in which the charging synchronization is avoided, which we attain by setting $C^S = 520$ \euro/MWh$^2$. Those cases are respectively denoted as \textit{naive-ch}, \textit{sync}, and \textit{non-sync}. Note that, in the former case, i.e. \textit{naive-ch}, each EV will be charged to its required maximum energy as soon as it is available, thus neglecting the dependence of the charging power on the price; whereas, the latter cases \textit{sync} and \textit{non-sync} are driven by the cost minimization of the EVs' aggregator wherein the electricity prices are accounted for. As an example, Fig. \ref{fig:power_sync} shows the EV-fleet charging power of a certain day for the three cases along with the electricity prices. As can be seen, the choice of $C^S \neq 0$ is a simple albeit convenient way to avoid the undesirable charging synchronization by smoothing the aggregate power. In addition, we can observe that the charging pattern of the \textit{naive-ch} case is independent of the prices. 

\begin{figure}[h]
\centerline{\includegraphics[scale=0.52]{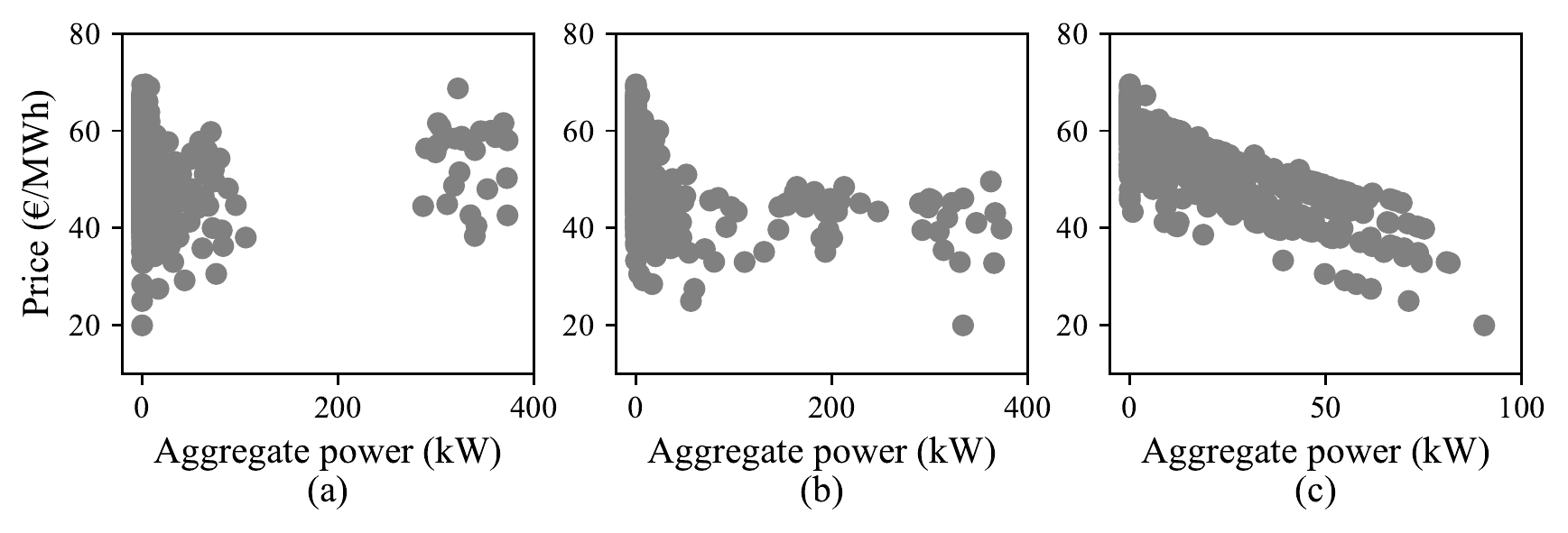}}
\vspace{-12pt}
\caption{ Power versus price for cases (a) \textit{naive-ch}, (b) \textit{sync}, and (c) \textit{non-sync}.}
\label{fig:power_price_g2v}
\end{figure}

The sizes of the training, validation, and test sets are 672 h, 168 h, and 168 h, in that order. Fig. \ref{fig:power_price_g2v} represents the hourly electricity price versus the corresponding charging power for all periods of the $\Omega^{tr}$ for the cases mentioned above. As can be seen, the aggregate power of the \textit{non-sync} case depends linearly on the price, unlike the \textit{naive-ch} and \textit{sync} cases. For the \textit{naive-ch} case, we consider 17 regressors, namely the charging power and the total number of EVs available for the six periods previous to time $t$, i.e., $p_{t-l}$ and $\sum_{v} \varsigma_{v,t-l}$, $\forall l=1...6$, and 5 binary-valued categorical variables to indicate the hour of the day. For the cases \textit{sync} and \textit{non-sync}, we consider 12 regressors, namely the electricity price and the charging power for the six periods previous to time $t$, i.e., $\lambda_{t-l}$ and $p_{t-l}$, $\forall l=1...6$.  We also assume six energy blocks in total. Finally, hyper-parameter $H$ ranges in the interval $[0.5, 1.0)$ with 0.01 steps, $M$ ranges in the interval $[0.0001, 0.0024]$ with 0.0001 steps, and $\gamma = \{0.1, 0.01\}$. For the case \textit{sync}, the proposed approach \textit{kio} takes on average 12.6 s, 2.6 s, and 31.3 s to run each \textit{feasibility} problem, \textit{optimality} problem, and all the forward problems for the $\Omega^{v}$, in that order. The computing times are of the same order of magnitude for the other cases. It should be noted that those computing times would be even suitable for an hour-ahead forecasting if the grid search technique were parallelized.   

The optimal hyper-parameters for all models and cases are given in Table \ref{tab:hyper_parameters_g2v}. The information given in this table is quite valuable and we can make two main remarks. First, cases \textit{sync} and \textit{non-sync} are price-driven and thus their optimal values of parameter $H^*$ are very high ($0.82$ and $0.94$ respectively) compared to the optimal value ($H^* = 0.64$) for the case \textit{naive-ch}, which is insensitive to the prices. In other words, the power bounds for the former cases are wider than for the latter one. Therefore, the \textit{optimality problem}, which is used to estimate the marginal utility, plays a major role to learn the aggregate response of the EV fleet for the price-driven cases. This is expected as the marginal utilities encode the impact of the current electricity price on the aggregate power of the EV fleet. Second, it should be noted that the optimal values of $H^*$ for the models \textit{kio} and \textit{lio} are quite similar, except for the case \textit{naive-ch}, for which \textit{lio} is unable to identify the insensitiveness of the aggregate power to the price.     

\begin{table}[h!]
\caption{Optimal Values of the Hyper-Parameters}
\label{tab:hyper_parameters_g2v}
\centering
\begin{tabular}{ccccc}
\hline
\multirow{1}{*}{Case} &  \textit{kio} &  \textit{krr} &  \textit{svr} &  \textit{lio} \\
\hline
\multirow{3}{*}{\textit{naive-ch}} & $H^*=0.64$ & $\delta^*=0.01$ & $C^*=100$ & $H^*=0.91$ \\
& $M^*=0.0002$ & $\gamma^*=0.1$ & $\gamma=0.01$ & \\
& $\gamma^*=0.1$ & & & \\
\hline
\multirow{3}{*}{\textit{sync}} & $H^*=0.82$ & $\delta^*=0.1$ & $C^*=10$ & $H^*=0.89$ \\
&  $M^*=0.0001$ & $\gamma^*=0.1$ & $\gamma^*=0.1$ &  \\
& $\gamma^*=0.1$ &  & & \\
\hline
\multirow{3}{*}{\textit{non-sync}} & $H^*=0.94$ & $\delta^*=0.1$ & $C^*=1$ & $H^*=0.94$ \\
& $M^*=0.002$ & $\gamma^*=0.1$ & $\gamma=0.1$ & \\
& $\gamma^*=0.01$ & & & \\
\hline
\end{tabular}
\end{table}

\begin{table}[h!]
\caption{Error Metrics -- Cases without V2G Services (\rm{kW})}
\label{tab:error_metrics_g2v}
\centering
\begin{tabular}{ccccccc}
\hline
\multirow{2}{*}{Model} & \multicolumn{2}{c}{\textit{naive-ch}}  &  \multicolumn{2}{c}{\textit{sync}}  &  \multicolumn{2}{c}{\textit{non-sync}} \\
\cline{2-7}
 & RMSE &  MAE&  RMSE &  MAE &  RMSE &  MAE \\
\hline
\textit{kio}  & 8.6 & 3.7 &  35.2 & 13.3 & 5.5 &  3.8 \\
\textit{krr}  & 9.0 & 3.5 &  35.5 & 15.7 & 7.4 & 5.2 \\
\textit{svr}  & 10.4 & 5.7&  41.7 & 14.7 & 7.6 & 5.0  \\
\textit{lio}  & 16.8 & 6.4 &  59.3 & 23.0 & 5.9 & 3.9 \\
\textit{h-naive} & 90.3 & 29.3  &  72.7 & 25.3 & 11.3 & 7.1 \\
\textit{d-naive} & 13.2 & 4.8 &  64.8 & 22.3 & 17.3 & 13.3  \\
\textit{w-naive} & 10.8 & 4.6 &  49.1 & 15.7 & 13.0 & 9.1 \\
\hline
\end{tabular}
\end{table}

\begin{figure}[h]
\centerline{\includegraphics[scale=0.6]{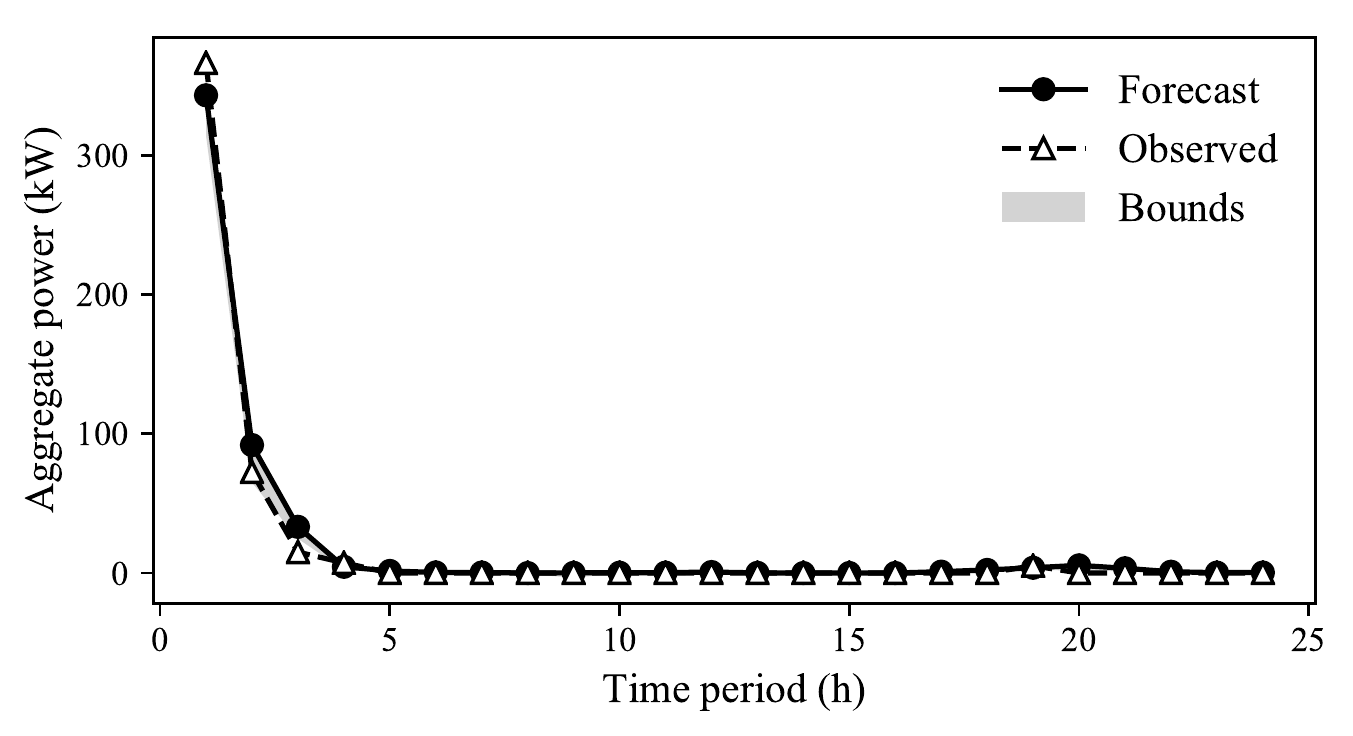}}
\vspace{-10pt}
\caption{Estimated power bounds as well as forecast and observed power for case \textit{naive-ch}.}
\label{fig:case_g2v_naive}
\end{figure}

The error metrics of the test set for all models are compared in Table \ref{tab:error_metrics_g2v} for the three cases. In the \textit{naive-ch} case, the least RMSE is obtained with the proposed model \textit{kio} with an error reduction of 4.4\% and 17.3\% compared to \textit{krr} and \textit{svr}. In the \textit{sync} case, the proposed model \textit{kio} achieves 28.3\% reduction in RMSE and 15.3\% reduction in MAE compared to the \textit{w-naive}, which provides the best performance among the naive models. As expected, we can also observe that the \textit{kio} outperforms \textit{lio} by reducing RMSE and MAE by 40.6\% and 42.2\% since \textit{kio} is able to capture the nonlinear relations between the EV-fleet power and the electricity price shown in Fig. \ref{fig:power_sync}. Finally, the performance of \textit{kio} is comparable to the performance of other machine-learning techniques such as \textit{krr} or \textit{svr}. In the \textit{non-sync} case, the aggregator behaves as a price-responsive EV fleet with a linear dependence and thus both \textit{kio} and \textit{lio} models achieve the least errors in the $\Omega^{test}$ compared to the other benchmarks. Note also that, in this case, the \textit{h-naive} is the one with the least error among the naive models. However, the RMSE of the \textit{kio} is decreased by 51.3\%, 25.7\%, and 27.6\% with respect to the one attained with the models \textit{h-naive}, \textit{krr}, and \textit{svr}, in that order. Overall, the \textit{kio} model is characterized for being versatile since it makes good predictions under any pattern of the EV-fleet power with the price.  

\begin{figure}[h]
\centerline{\includegraphics[scale=0.6]{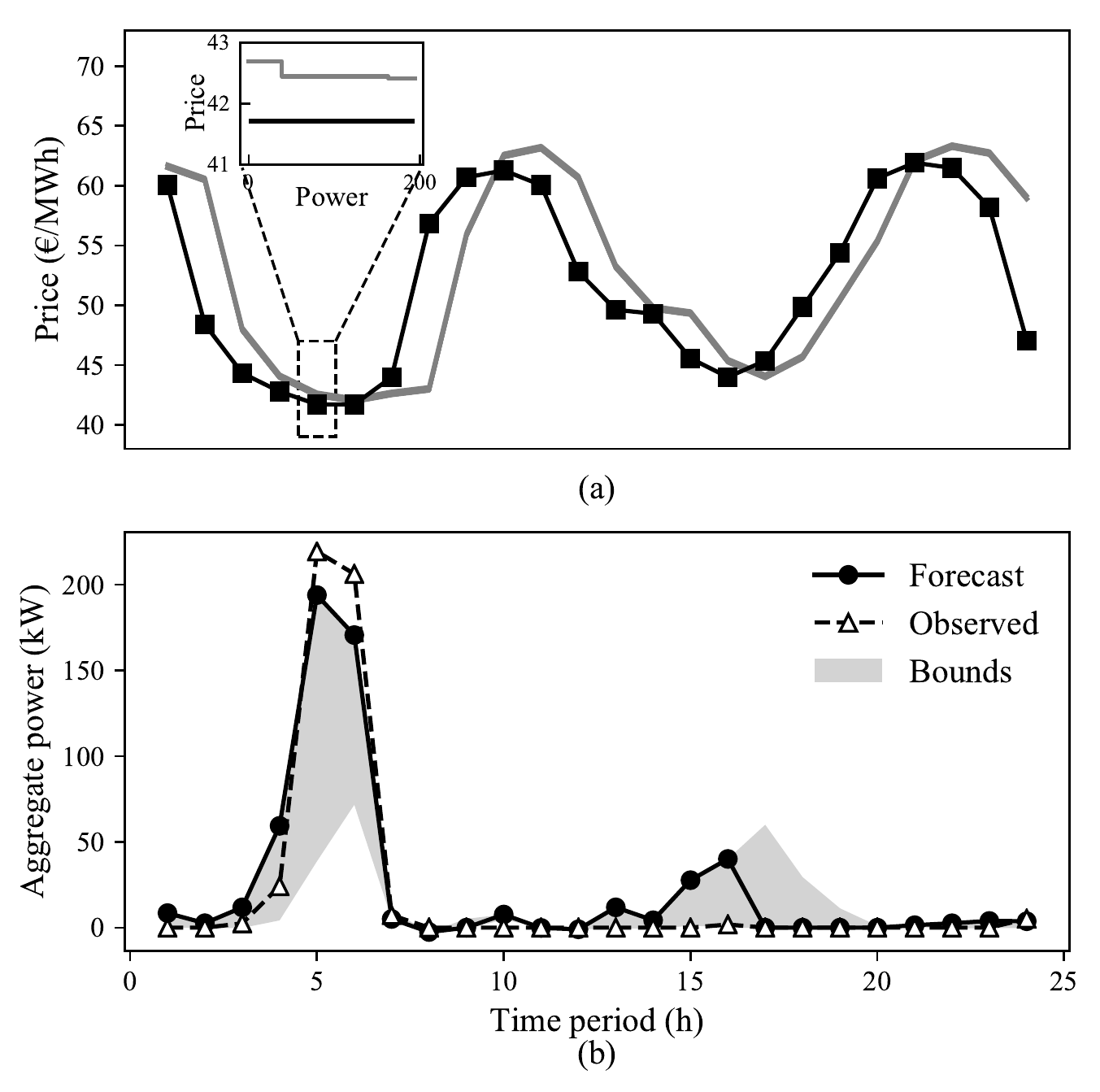}}
\vspace{-10pt}
\caption{Results for case \textit{sync}: (a) Estimated marginal utility price per block (in grey) and electricity price (in black) and (b) estimated power bounds as well as forecast and observed power. Note that the inset plot represents the bid price function and the corresponding electricity price of hour 5.}
\label{fig:case_g2v_a0}
\end{figure}

\begin{figure}[h]
\centerline{\includegraphics[scale=0.6]{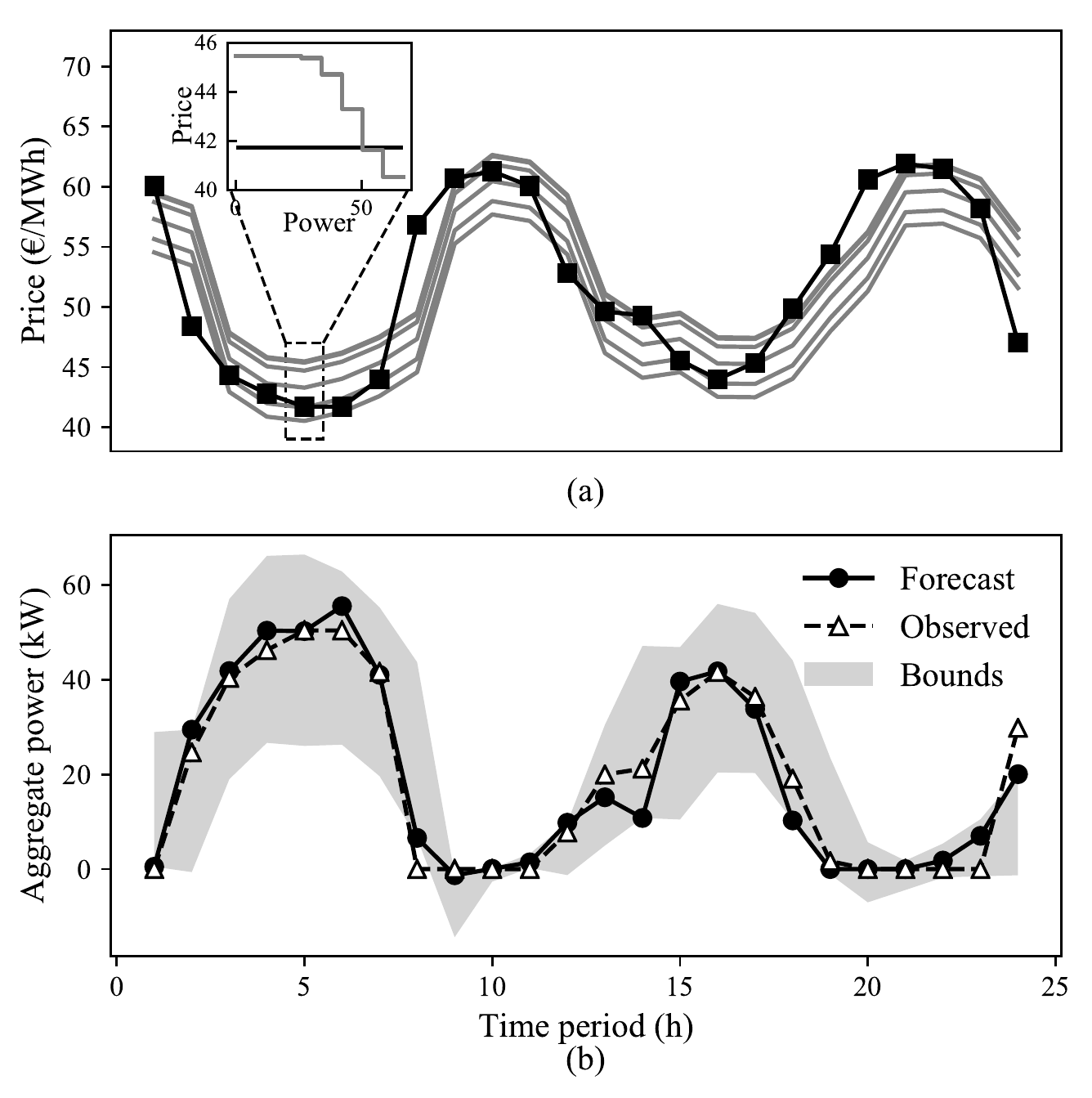}}
\vspace{-10pt}
\caption{Results for case \textit{non-sync}: (a) Estimated marginal utility price per block (in grey) and electricity price (in black) and (b) estimated power bounds as well as forecast and observed power. Note that the inset plot represents the bid price function and the corresponding electricity price of hour 5.}
\label{fig:case_g2v_anot0}
\end{figure}

Apart from the improvement in terms of RMSE and MAE of the \textit{kio} against the rest of the models to learn the EV-fleet power, the proposed approach is able to provide a bid curve, as imposed by rules in electricity markets \citep{omie}. Figures \ref{fig:case_g2v_naive}--\ref{fig:case_g2v_anot0} show the results for cases \textit{naive-ch}, \textit{sync}, and \textit{non-sync}, respectively. In Fig. \ref{fig:case_g2v_a0}.(a) and \ref{fig:case_g2v_anot0}.(a), we show the estimated marginal utilities for the six blocks for each hour of the first day of the $\Omega^{test}$ and for the cases \textit{sync} and \textit{non-sync}. In those figures, we also show the decreasing bid curves at hour 5 in the inset plots, which are also presented in Tables \ref{tab:bid_curve_sync}--\ref{tab:bid_curve_nonsync}. Correspondingly, Fig. \ref{fig:case_g2v_naive}, \ref{fig:case_g2v_a0}.(b), \ref{fig:case_g2v_anot0}.(b) depict the estimated bounds as well as the forecast and observed EV-fleet power for such a day. 

In the \textit{naive-ch} case, the \textit{kio} provides coincident power bounds, as illustrated in Fig. \ref{fig:case_g2v_naive}, which means that the \textit{optimality problem} (i.e. the marginal utility estimation problem, which captures the price effect) is useless and thus the aggregate charging power can be directly explained by estimating the bounds. In Fig. \ref{fig:case_g2v_a0}.(a) and \ref{fig:case_g2v_anot0}.(a), we can observe that the \textit{kio} model identifies whether the EV-fleet power is price-responsive or not by assigning different values to the marginal utility for each block. On the one hand, in Fig. \ref{fig:case_g2v_a0}.(a), the blockwise marginal utilities are almost identical at any time period, thus suggesting an almost all-or-nothing price response of the EV fleet for the \textit{sync} case. In this case, the power bounds are basically shaping the EV-fleet charging forecast. On the other hand, for the \textit{non-sync} case, the bounds are generally wider than those obtained for the \textit{sync} case (see Fig. \ref{fig:case_g2v_anot0}.(b)). The marginal utility is thus shaping the aggregate power forecast since the \textit{kio} model gives rise to a wider range of marginal utility values at any time period, as can be observed in Fig. \ref{fig:case_g2v_anot0}.(a). In short, unlike any other forecasting tool, we gain interpretability with the proposed IO approach \textit{kio} due to two aspects: (i) the width of the bounds, which sheds light on the price-responsiveness of the EV fleet; and (ii) the derivation of a bid curve when there exists a dependence of the EV-fleet power on the price, as can be seen in the inset plots of Figs. \ref{fig:case_g2v_a0}.(a)--\ref{fig:case_g2v_anot0}.(a) and Tables \ref{tab:bid_curve_sync}--\ref{tab:bid_curve_nonsync}. 

\begin{table}[h]
\caption{Bid Curve at Hour 5 -- Case \textit{sync}}
\label{tab:bid_curve_sync}
\centering
\begin{tabular}{ccccccc}
\hline
Block & 1 & 2 & 3 & 4 & 5 & 6\\
\hline
Marginal utility (€/MWh) &42.7	&42.4&42.4&42.4&42.4&42.4\\
\hline
Power block (kW) &38.7	&31.1&31.1&31.1&31.1&31.1\\
\hline
\end{tabular}
\end{table}

\begin{table}[h]
\caption{Bid Curve at Hour 5 -- Case \textit{non-sync}}
\label{tab:bid_curve_nonsync}
\centering
\begin{tabular}{ccccccc}
\hline
Block & 1 & 2 & 3 & 4 & 5 & 6\\
\hline
Marginal utility (€/MWh) &45.5 &45.4 &44.7 &43.3 &41.6 &40.5\\
\hline
Power block (kW) &26.0 &8.1 &8.1 &8.1 &8.1 &8.1\\
\hline
\end{tabular}
\end{table}

\subsection{Forecast Results with V2G Services}
\label{sec:results_v2g}
We now assume that EVs may enable their V2G capabilities (i.e. $B_v^d \neq 0$ in the model \eqref{sim_of}--\eqref{sim_eq8}) and we compare the results for two cases: (i) a highly-synchronized power case when $C^S = 0$; and (ii) a case in which the power synchronization is avoided when $C^S = 52$ \euro/MWh$^2$. Those cases are denoted as \textit{sync} and \textit{non-sync}. The problem setup is identical to that explained in Section \ref{sec:results_g2v}. Table \ref{tab:error_metrics_v2g} provides the error metrics on the $\Omega^{test}$ for all models. As can be seen, \textit{kio} clearly outperforms by far the \textit{lio} and naive models for both cases. Notwithstanding, the performance of \textit{lio} in terms of error is closer to the proposed approach for the \textit{non-sync} case because the EV-fleet power is more price-responsive. Also, the performance of \textit{kio} is similar to the machine-learning techniques \textit{krr} and \textit{svr} in the case \textit{sync}; and the RMSE (MAE) decreases by 4.8\% and 5.9\%  (11.4\% and  6.7\%) compared to \textit{krr} and \textit{svr}, respectively, in the case \textit{non-sync}.

\begin{table}[h!]
\caption{Error Metrics -- Cases with V2G Services (\rm{kW})}
\label{tab:error_metrics_v2g}
\centering
\begin{tabular}{ccccc}
\hline
\multirow{2}{*}{Model} &  \multicolumn{2}{c}{\textit{sync}}  &  \multicolumn{2}{c}{\textit{non-sync}} \\
\cline{2-5}
 & RMSE &  MAE&  RMSE &  MAE  \\
\hline
\textit{kio}  &  148.6 & 94.3 & 33.5 & 20.9\\
\textit{krr}  &  146.9 & 108.4 & 35.2 & 23.6\\
\textit{svr}  &   147.1 & 92.4 & 35.6 & 22.4\\
\textit{lio}  &  172.1 & 120.0 & 36.2 & 23.7\\
\textit{h-naive} &  235.4 & 142.2 & 49.5 & 30.0\\
\textit{d-naive} &  261.8 & 162.5 & 71.1 & 50.2\\
\textit{w-naive} & 199.5 & 112.3 & 60.4 & 37.7\\
\hline
\end{tabular}
\end{table}

\section{Conclusions}
\label{sec:conclusion}
This paper proposes a data-driven two-step estimation procedure relying on two main concepts: inverse optimization and kernel regression. This novel approach allows to capture the nonlinear relationship between an aggregate price-response and the associated explanatory variables, while deriving a bid/offer curve, as imposed by rules in electricity markets. We apply such a framework to learn the aggregate price-response of an EV fleet. The proposed approach attains a better performance (around 20\%--40\% error reduction) than naive or linear models. Moreover, it achieves a similar or better (depending on the case) performance than state-of-the-art machine-learning techniques such as support vector regression or kernel-ridge regression. Overall, the proposed approach is versatile since its performance is good regardless of the price-power relation. Very interestingly, besides, it increases the degree of interpretability of the prediction model compared to existing approaches in the literature since a bid/offer curve can be readily derived.

\section{Acknowledgements}
This project has received funding in part by the Spanish Ministry of Economy, Industry, and Competitiveness through project ENE2017-83775-P; in part by the European Research Council (ERC) under the European Union's Horizon 2020 research and innovation programme (grant agreement No 755705); and in part by Fundaci\'on Iberdrola Espa\~na 2018. The authors thankfully acknowledge the computer resources, technical expertise and assistance provided by the SCBI (Supercomputing and Bioinformatics) center of the University of Malaga.

\vspace{-0.5cm}

\appendix
\section{Aggregator of Electric Vehicles}
\label{sec:simulator}
\vspace{-0.3cm}
To simulate the behavior of a pool of EVs, i.e., its aggregate power, we assume an aggregator of EVs in residential districts who aims to minimize their total costs. This can be mathematically expressed as:
\begin{align}
% Object Function 1
&\min_{\Xi^{ev}} \sum_{t \in \mathcal{T}}  \hspace{-2pt}\Bigg( \lambda_{t} \Delta t p_t + \hspace{-2pt} \sum_{v \in \mathcal{V}} \hspace{-2pt}  \Big(  C_{v,t}^{D} \hspace{-1pt} + \hspace{-1pt} C^{P} s_{v,t} \Big) \hspace{-1pt} + \hspace{-1pt} C^S \Delta t^2 p_t^2 \Bigg) \label{sim_of} \\
&\text{subject to:}\notag\\
& p_t = \sum_{v \in \mathcal{V}} \left( c_{v,t} - d_{v,t} \right), \quad \forall t \in \mathcal{T} \label{sim_eq1}\\
& {soc}_{v,t} = {soc}_{v,t-1} + \Delta t \left( \eta^{c}_{v} c_{v,t} - \frac{d_{v,t}}{\eta^{d}} \right) - \chi_{v, t} + s_{v,t} \quad \forall v \in \mathcal{V}, t \in \mathcal{T} \label{sim_eq2}\\
& 0 \leq c_{v,t} \leq B^c_{v} \varsigma_{v,t} , \quad \forall v \in \mathcal{V}, t \in \mathcal{T} \label{sim_eq3}\\
& 0 \leq d_{v,t} \leq B^d_{v} \varsigma_{v,t} , \quad \forall v \in \mathcal{V}, t \in \mathcal{T} \label{sim_eq4}\\
& \underline{SOC}_{v,t} \leq {soc}_{v,t} \leq \overline{SOC}_{v,t}, \quad \forall v \in \mathcal{V}, t \in \mathcal{T} \label{sim_eq5}\\
& {soc}_{v,N_T} = {soc}_{v, 0}, \quad \forall v \in \mathcal{V} \label{sim_eq6}\\
& C^{D}_{v,t} = A_{v,t} + F_v d_{v,t}, \quad \forall v \in \mathcal{V}, t \in \mathcal{T} \label{sim_eq7}\\
&  s_{v,t} \geq 0, \quad \forall v \in \mathcal{V}, t \in \mathcal{T}, \label{sim_eq8}
\end{align}
\noindent where the set of decision variables $\Xi^{ev} = \{c_{v,t}, C_{v,t}^{D}, d_{v,t}, p_t, s_{v,t}, {soc}_{v,t} \}$, $\mathcal{V}$ is the set of EVs in the fleet, $\mathcal{T}$ is the set of time periods. The variable $p_t$ represents the power the aggregator buys in the electricity market whereas the variables $c_{v,t}$ and $d_{v,t}$ represent the charging power from and discharging power to the grid of EV $v$ in period $t$. The variable $C^{D}_{v,t}$ represents the cost of battery degradation due to motion and charging/discharging cycle of EV $v$ in period $t$. The variables $s_{v,t}$ act as a load shedding term when the energy balance of the EVs cannot be satisfied. Finally, ${soc}_{v,t}$ is the state of charge of the battery of EV $v$ in period $t$. In addition, $\lambda_{t}$ is the electricity price in period $t$; $\Delta t$ is the time step; $C^P$ is the load shedding cost; $C^S$ is a penalty cost to avoid power synchronization; $\eta^{c(d)}_{v}$ is the charging (discharging) efficiency for the EV $v$; $\chi_{v,t}$ represents the energy required for transportation of each EV throughout the time horizon; $B^c_{v}$ and $B^d_{v}$ are the maximum charging and discharging power of EV $v$, respectively; $\varsigma_{v,t}$ represents the availability of the EV $v$ in period $t$; $\underline{SOC}_{v,t}$ and $\overline{SOC}_{v,t}$ are the minimum and maximum limits of the energy state of charge of EV $v$ in period $t$; $N_T$ is the number of time periods; $F_{v}$ is the degradation cost per kW due to charging-discharging cycles and it depends on the battery cost of EV $v$; and $A_{v,t}$ is the degradation cost due to motion of EV $v$ in time period $t$.

The problem \eqref{sim_of}--\eqref{sim_eq8} aims to minimize the total costs as given in \eqref{sim_of}, which comprise four terms: (i) the operational costs due to charging from and discharging to the grid, (ii) the degradation costs of the vehicles' batteries, (iii) the load shedding costs when the equation associated with the energy state-of-charge evolution is violated, and (iv) the penalty costs to avoid power synchronization that may lead to overloads in the distribution network \citep{sarker2016electric}. Constraints \eqref{sim_eq1} relate the power bought in the electricity market with the charging and discharging power. Constraints \eqref{sim_eq2} model the energy state of charge evolution while taking into account the energy required for transportation. Expressions \eqref{sim_eq3} and \eqref{sim_eq4} impose the lower and upper bounds for the charging and discharging power, in that order. Constraints \eqref{sim_eq5} set the lower and upper bounds for the energy state of charge of the EVs. Expressions \eqref{sim_eq6} enforce boundary conditions on the energy
state-of-charge of the EVs. Expressions \eqref{sim_eq7} model the battery degradation costs based on the motion status and the discharging energy, as described in \cite{ortega2014optimal}. Finally, constraints \eqref{sim_eq8} define the non-negativity character of the variable $s_{v,t}$.

\vspace{-0.5cm}
%\section*{References}

\bibliography{EV_invfor}

\end{document}